\documentclass[a4paper,12pt]{revtex4}

\pdfoutput=1

\usepackage[english]{babel}
\usepackage{epsfig,graphicx,color}

\begin{document}

\title{Size dependent oscillator strength and quantum efficiency of CdSe quantum dots determined by controlling the local density of states}

\author{M. D. Leistikow$^1$, J. Johansen$^2$, A. J. Kettelarij$^1$, P. Lodahl$^2$, and W. L. Vos$^{1, 3}$}
\affiliation{1) Center for Nanophotonics, FOM Institute for Atomic
and Molecular Physics (AMOLF), Kruislaan 407
1098 SJ, Amsterdam, The Netherlands\\
2) DTU Fotonik, Department of Photonics Engineering, Technical
University of Denmark, DK-2800 Kgs. Lyngby, Denmark \\
3) Complex Photonic Systems (COPS), Faculty of Science and
Technology and MESA$^+$ Institute for Nanotechnology, University of
Twente, 7500 AE Enschede, The Netherlands}

\homepage{www.photonicbandgaps.com}


\date{\today}

\begin{abstract}

We study experimentally time-resolved emission of CdSe quantum dots
in an environment with a controlled local density of states (LDOS).
The decay rate is measured versus frequency and as a function of
distance to a mirror. We observe a linear relation between the decay
rate and the LDOS, allowing us to determine the size-dependent
quantum efficiency and oscillator strength. We find that the quantum
efficiency decreases with increasing emission energy mostly due to
an increase in nonradiative decay. For the first time, we manage to
obtain the oscillator strength of the important class of CdSe
quantum dots. The oscillator strength varies weakly with frequency
in agreement with behavior of quantum dots in the strong confinement
limit. Surprisingly, the measured absolute values are a factor of 5
below theoretically calculated values. Our results are relevant for
applications of CdSe quantum dots in spontaneous emission control
and cavity quantum electrodynamics.
\end{abstract}

\maketitle


\section{Introduction}

Control over spontaneous emission is important for many applications
in nanophotonics, such as efficient miniature lasers and LEDs
\cite{yablonovitch, park}, efficient solar energy collection
\cite{gratzel}, and even biophotonics \cite{blum}. Increasing
attention has been given to all solid state cavity quantum
electrodynamics (QED) experiments \cite{yoshie, reithmaier,
lodahl,peter}. For spontaneous emission control the oscillator
strength plays a crucial role. The oscillator strength gauges the
strength of the interaction of a light source with the light field.
The larger the oscillator strength is, the stronger is the
interaction between the source and the light field, and in cavity
QED between source and cavity field.

As light sources in nanophotonics, quantum dots are becoming
increasingly popular. Quantum dots are semiconductor nanocrystals
with sizes smaller than the exciton Bohr radius. Due to their small
size, quantum dots have discrete energy levels \cite{brus}. CdSe
colloidal quantum dots in particular have generated enormous
interest in recent years because of the tunability of their emission
energy over the entire visible range with particle diameter
\cite{efros}. Surprisingly no measurements have been done of the
emission oscillator strength of these quantum dots, while this is
highly important to interpret cavity QED experiments
\cite{lethomas}. The oscillator strength has been investigated only
qualitatively using absorption measurements \cite{schmelz, striolo,
leatherdale}. However, the accuracy of these measurements are
limited due to the strong blinking behavior of CdSe quantum dots,
i.e. intermittency in the emission of photons. The oscillator
strength determined from absorption is not relevant to emission
experiments since the quantum dots in the off state do absorb while
they do not contribute to the emission.

In this article we present quantitative measurements of the
oscillator strength and quantum efficiency of colloidal CdSe quantum
dots as a function of emission energy and dot diameter since the
emission energy and diameter are uniquely related \cite{efros}. The
oscillator strength of an emitter can be determined by placing it
close to an interface. The emission rate will be affected by
emission which is reflected at the interface and leads to a
controlled modification of the local density of states (LDOS)
allowing us to separate radiative and nonradiative decay rate
components. This technique has been pioneered by Drexhage for dye
molecules \cite{drexhage} and used to determine quantum efficiency
of Si nanocrystals \cite{walters}, erbium ions \cite{snoeks},
epitaxially grown InAs quantum dots \cite{johansen} and colloidal
CdSe quantum dots \cite{brokmann, zhang}. Recently it has been found
that the emission oscillator strength can also be determined with
this technique \cite{johansen}. Here, we place CdSe quantum dots on
different distances near a silver interface to quantitatively
determine the oscillator strength as a function of emission energy.


\section{Experimental Methods}

\subsection{Sample fabrication}

\begin{figure}[!tbp]
\includegraphics[width=10cm]{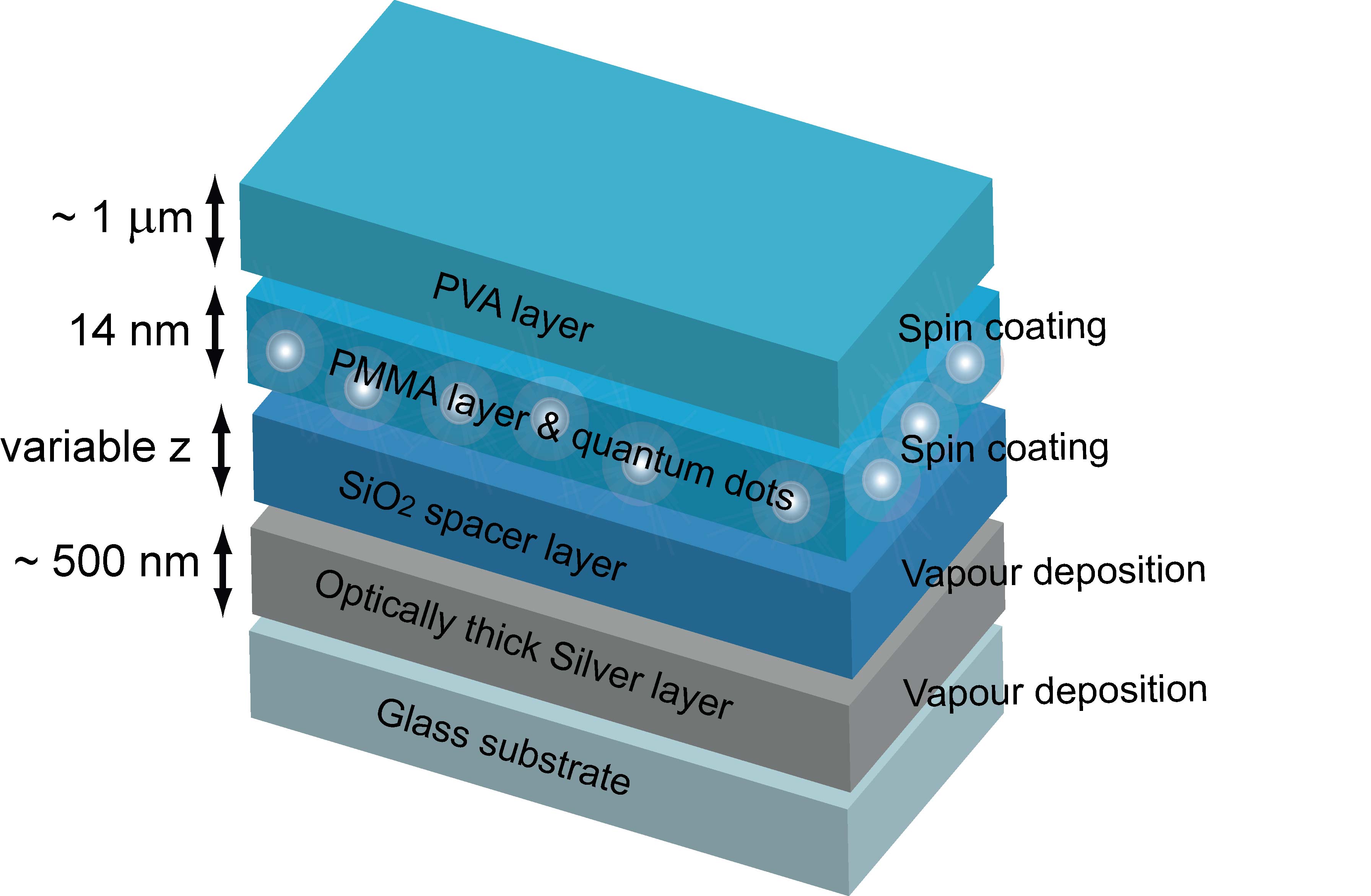}
\caption{Schematic cross-section of the sample used in the
measurements. The different layers of the sample shown together with
corresponding thickness and fabrication technique.}\label{sample}
\end{figure}

The planar samples with controlable LDOS consist of a glass
substrate of 24 by 24 mm on which a stack of 4 different layers is
made, as shown in figure \ref{sample}. 1) The first layer is an
optically thick 500 nm layer of silver which is deposited with vapor
deposition. 2) Next a layer of SiO$_2$ is evaporated onto the
silver. The SiO$_2$ layer has a refractive index of $1.55$ $\pm$
$0.01$ at a wavelength of 600 nm as determined by ellipsometry. The
thickness of the SiO$_2$ layer is varied to control the distance z
that the quantum dots have to the silver interface. 3) On top of the
SiO$_2$ layer, a very thin layer of polymethyl methacrylate (PMMA)
is spincoated that contains the CdSe quantum dots. This layer is
$\Delta z = 14 \pm 5$ nm thick, determined by profilometry. PMMA has
a refractive index of $1.49 \pm 0.01$. 4) On top of the PMMA layer a
thick $\sim 1\mu m$ layer of polyvinyl alcohol (PVA) is spincoated
to avoid reflections from a PMMA/air interface. The PVA is 9.4 \% by
weight dissolved in a mixture of water and ethanol. Since the PMMA
and quantum dots do not dissolve in water and ethanol, the PMMA
layer stays intact. PVA has a refractive index of $1.50 \pm 0.01$.
All parameters are summarized in Table \ref{layerstab}.

\begin{table}[ht]
\caption{Layer properties} 
\centering 
\begin{tabular}{c c c c} 
\hline\hline 
Layer & Thickness (nm) & Refractive index & Fabrication method \\ [0.5ex] 
\hline 
1) Silver & 500 & $0.27+4.18 i$ & vapor deposition \\ 
2) SiO$_2$ & variable z & 1.55 & vapor deposition \\
3) PMMA + CdSe quantum dots & 14 $\pm$ 5 & 1.49 & spincoating \\
4) PVA & $\sim 1000$ & 1.50 & spincoating \\[1ex] 
\hline 
\end{tabular}
\label{layerstab} 
\end{table}

\subsection{Quantum dots}
CdSe quantum dots with a ZnS shell are purchased from Evident
Technology (Fort Orange, emitting around 600 nm). The suspension
that is spincoated consists of toluene with 0.5 \% by weight 495,000
molecular weight PMMA and a quantum dot concentration of $1.21$
$10^{-6}$ mol/liter. The quantum dots have an estimated density of 1
per 2500 nm$^2$. The quantum dots are thus sufficiently dilute in
the PMMA layer to exclude energy transfer and reabsorption processes
between quantum dots. This was verified by measuring that the decay
rate was not influenced by laser power or changes in concentration
around the used concentration. The sample is contained in a nitrogen
purged chamber during measurements to prevent photo oxidation of the
quantum dots.

\subsection{Optical detection}

\begin{figure}[!tbp]
\includegraphics[width=10cm]{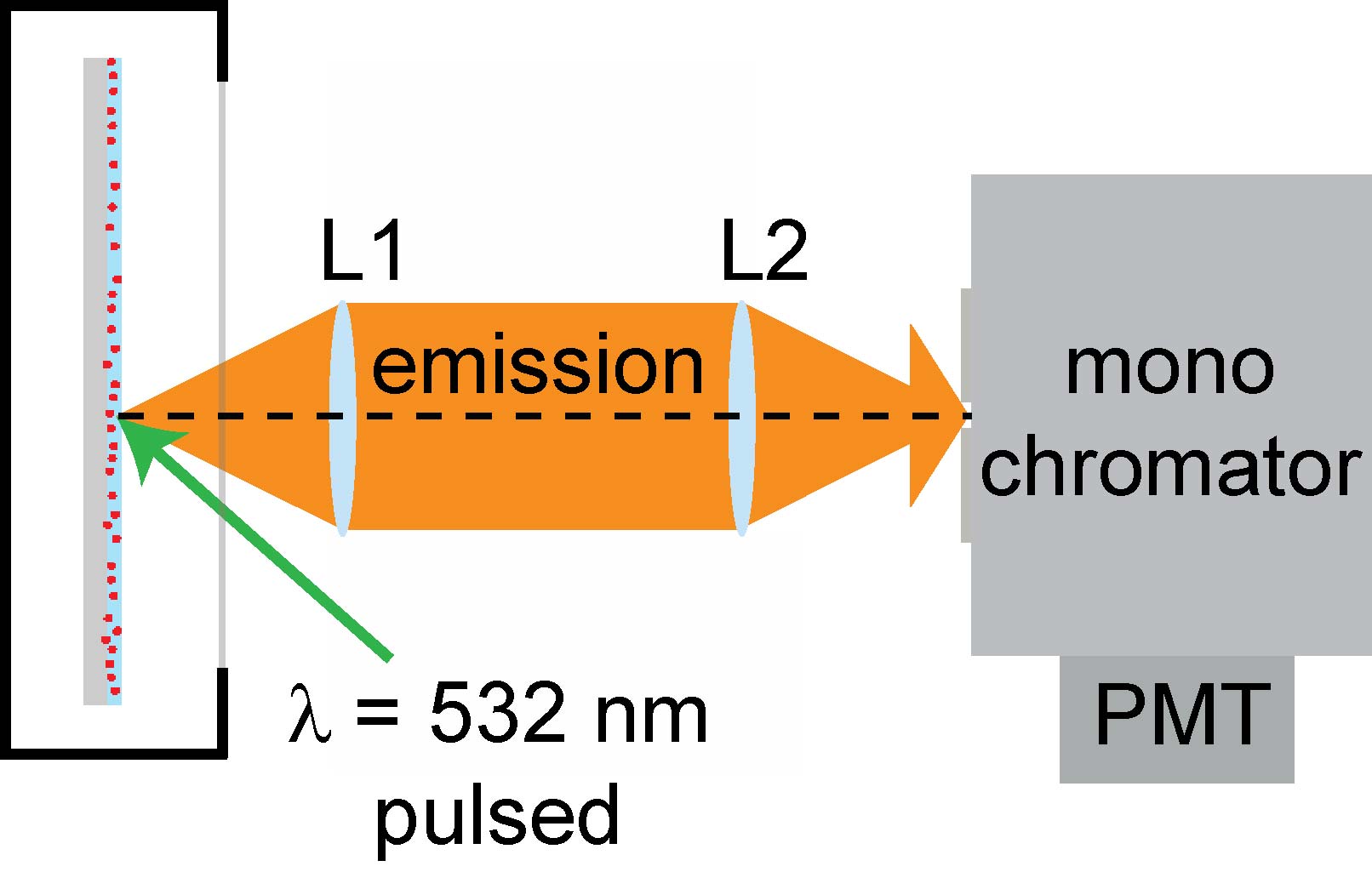}\caption{A schematic picture
of the experimental setup. Light from the laser excites the quantum
dots in a layered sample inside a nitrogen purged chamber. The
emitted light is collimated by a lens L1 with f=12 cm, focused by
lens L2 with f=10 cm on the entrance slit of a monochromator and
detected by the photomultiplier tube.}\label{setup}
\end{figure}

The optical set-up used in the experiments is schematically shown in
figure \ref{setup}. Light from a pulsed frequency doubled
Nd$^{3+}$:YAG laser (Time Bandwidth Cougar) with an emission
wavelength of 532 nm, repetition rate of 8.2 MHz and pulse widths of
11 ps is used. This light is guided into an optical fiber and
focused onto the sample by a lens with a focal length of 250 mm,
leading to a focus with a diameter of approximately 50 $\mu$m on the
sample.

The light emitted by the quantum dots is collected by a lens,
collimated and focused onto the slit of a prism monochromator (Carl
Leiss). The slit width is set to 400 $\mu$m giving a spectral
resolution $\Delta\lambda= 6$ nm, which is narrow compared to the
bandwidth of the LDOS changes. A Hamamatsu multichannel plate
photomultiplier tube is used as a photon counter. With this setup it
is possible to measure spectra by scanning the monochromator and to
measure decay curves of emitters at particular emission frequencies
by time correlated single photon counting \cite{Oconnor}. This
technique measures the time between the arrival of an emitted photon
(start) and the laser pulse (stop) with ps resolution. By repeating
such a measurement a histogram of the arrival times is made from
which a decay rate can be determined. The time resolution of the
set-up is 125 ps, given by the full width half maximum of the total
instrument response function that is shown in figure
\ref{decayinPMMA}. The instrument response function is much shorter
than the decay curve of CdSe quantum dots, with a typical decay time
of 16 ns in toluene. Therefore, deconvolution of the response
function is not performed before analyzing the data.

\subsection{Data interpretation}

The quantum dots in the polymer layer show a nonexponential decay,
probably caused by microscopic heterogeneity of the polymer
\cite{vallee}. Nonexponential behavior has previously been found for
CdSe quantum dots in PMMA by Fisher et al. \cite{fisher} even for
single quantum dots. To model the decay curve the data are fitted
with a distribution of decay rates as explained in ref.
\cite{vandriel2}. A function of the following form is used to model
the decay curve:

\begin{equation}
f(t)=\int_0^{\infty}\sigma(\gamma_{tot}) \exp(-\gamma_{tot} t)
\rm{d}\gamma_{tot}
\end{equation}

where the normalized distribution in decay rates is chosen to be
lognormal

\begin{equation}
\sigma(\gamma)=A~{\exp}{\Big[-\Big(\frac{ln(\gamma)-ln(\gamma_{mf})}{w}\Big)^2\Big]}
\end{equation}

The normalization factor A equals $A=[\gamma_{mf} w ~\sqrt\pi~
\exp(w^2/4)]^{-1}$. The two relevant adjustable parameters that can
be extracted from the model are the most frequent decay rate
$\gamma_{mf}$ which is the peak of the lognormal distribution and
$\Delta\gamma = 2 \gamma_{mf} \sinh(w)$ which is the $\frac{1}{e}$
width of the lognormal distribution.

Decay rates presented in this paper are an average of decay rates
found for at least three measurements performed on different
locations on a sample with a particular SiO$_2$ layer thickness. The
error in the decay rate is conservatively estimated to be $\pm$ 3
$\%$ which is the maximum difference found between measurements on
the same sample.

\newpage

\section{Results}

\subsection{Experimental results}
\begin{figure}[!tbp]
\includegraphics[width=10cm]{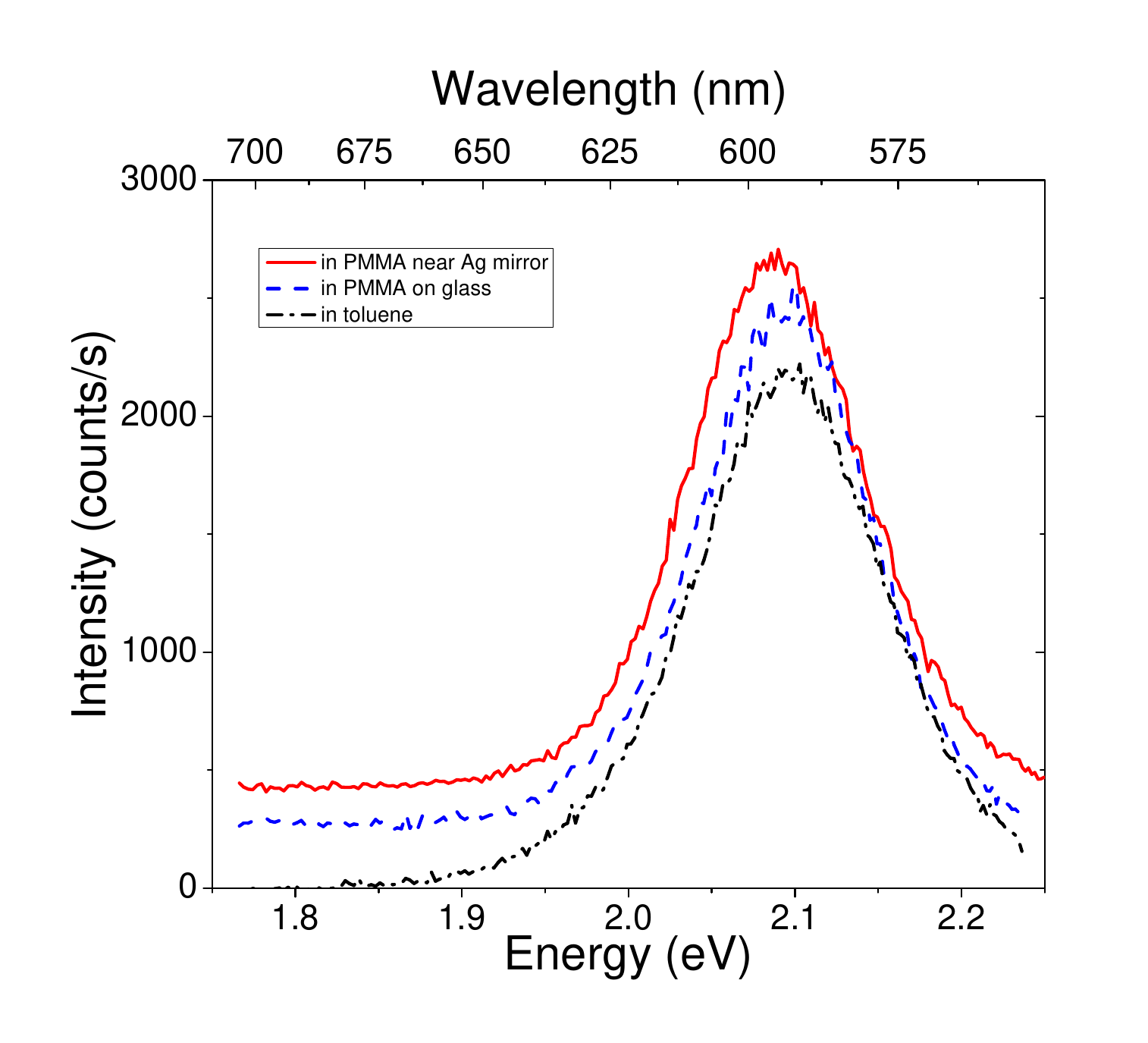}
\caption{(color) Emission spectra of CdSe quantum dots in toluene
suspension, in a planar sample without silver, and in a planar
sample with a silver mirror. The spectra are offset for clarity by
200 and 400 counts/s respectively. The spectrum in PMMA near the
mirror and in toluene are scaled to the spectrum in PMMA on glass by
a factor of 0.75.}\label{spectrumtoluene}
\end{figure}

In figure \ref{spectrumtoluene} the emission spectrum of CdSe
quantum dots is shown for the quantum dots in toluene, in a planar
sample without silver, and in a planar sample with a silver mirror.
The peak energies of all three spectra are identical within
experimental error. The width of the spectrum is caused by
inhomogeneous broadening due to size polydispersity of quantum dots
in the ensemble. The homogeneous spectral width of the individual
quantum dots is much narrower \cite{empedocles}. By selecting a
narrow emission energy window quantum dots of a particular diameter
are selected. Within experimental error there is no difference
between the width of the emission spectra in the different
environments, indicating that there is no spectral broadening due to
the polymer environment.

\begin{figure}[!tbp]
\includegraphics[width=10cm]{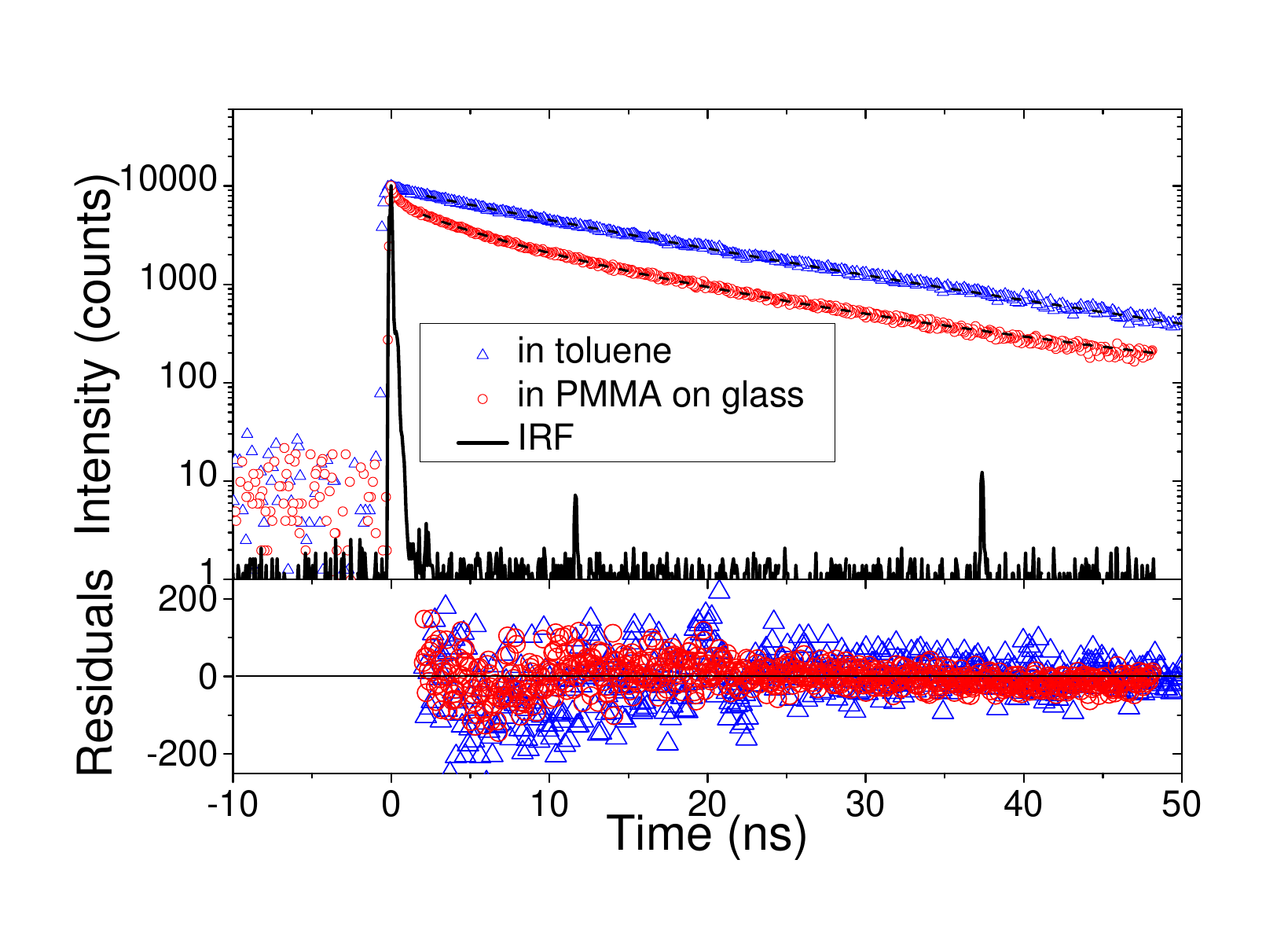}
\caption{(color) Decay curves of quantum dots at the emission peak
at 2.08 eV in PMMA on glass with a top layer of PVA (red circles)
and these quantum dots in toluene suspension (blue triangles). The
instrument response function (IRF) is indicated by the black line.
The peaks in the IRF near 12 and 36 ns are related to the pulse
picker of the laser. The decay curves are fitted with a lognormal
distribution of decay rates. Residuals are shown in the bottom
panel. }\label{decayinPMMA}
\end{figure}

In figure \ref{decayinPMMA} decay curves are shown at the emission
peak at 2.08 eV for an ensemble of quantum dots in toluene
suspension and in a planar layer without mirror. The quantum dots in
toluene show a single exponential decay as expected, giving a decay
rate $\gamma=0.061$ ns$^{-1} \pm 0.002$. Fitting the data with a
single exponential gives a value of 1.94 for the goodness of fit
$\chi_{red}^{2}$ indicative of a reasonable fit \cite{lakowicz}.

\begin{figure}[!tbp]
\includegraphics[width=10cm]{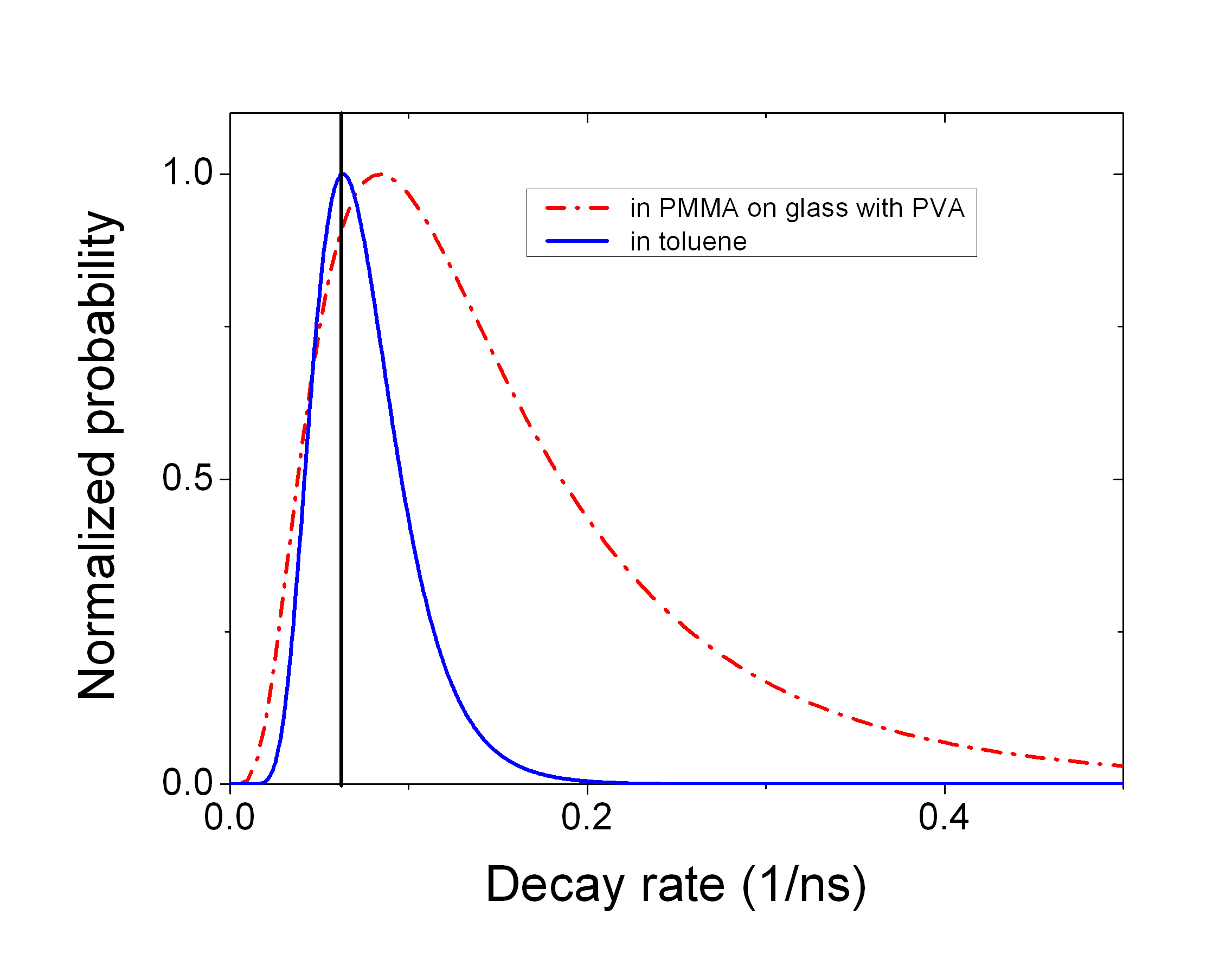}
\caption{(color) Lognormal distribution of decay rates of quantum
dots in a PMMA layer on glass with a PVA cover layer and for quantum
dots in toluene resulting from fits in figure \ref{decayinPMMA}. The
black line shows the delta function distribution for single
exponential fit. }\label{distributions}
\end{figure}

The lognormal distribution of decay rates can be fitted to the decay
curve of quantum dots inside PMMA and appears to be a good fit with
$\chi_{red}^{2}=1.49$. For the quantum dots inside the PMMA layer
$\gamma_{mf}=0.084$ ns$^{-1} \pm 0.002$. This decay rate indicates
the peak in the distribution. The decay of spontaneous emission from
quantum dots in toluene suspension can also be fitted with a
lognormal distribution of decay rates, giving $\chi_{red}^{2}=1.71$.
The distribution of decay rates in toluene is characterised by
$\gamma_{mf}$ = 0.063 ns$^{-1} \pm 0.002$ close to the value for the
decay rate $\gamma=0.061$ ns$^{-1} \pm 0.002$ found from a single
exponential decay. In figure \ref{distributions} the lognormal
distributions of decay rates are shown for the decay curve of
quantum dots in toluene and in the polymer layer. The distribution
of decay rates for quantum dots in polymer is much broader than the
distribution found for quantum dots in toluene. When a curve is
modeled with a single exponential decay the decay rate distribution
reduces to a delta function (indicated in black). The decay rate at
the peak of the distribution, the most frequent decay rate,
characterizes the decay in the measurement best as supported by the
fact that the $\gamma$ and $\gamma_{mf}$ for decay in toluene are
equal within experimental error. The most frequent decay rate will
be used in further analysis.

\begin{figure}[!tbp]
\includegraphics[width=10cm]{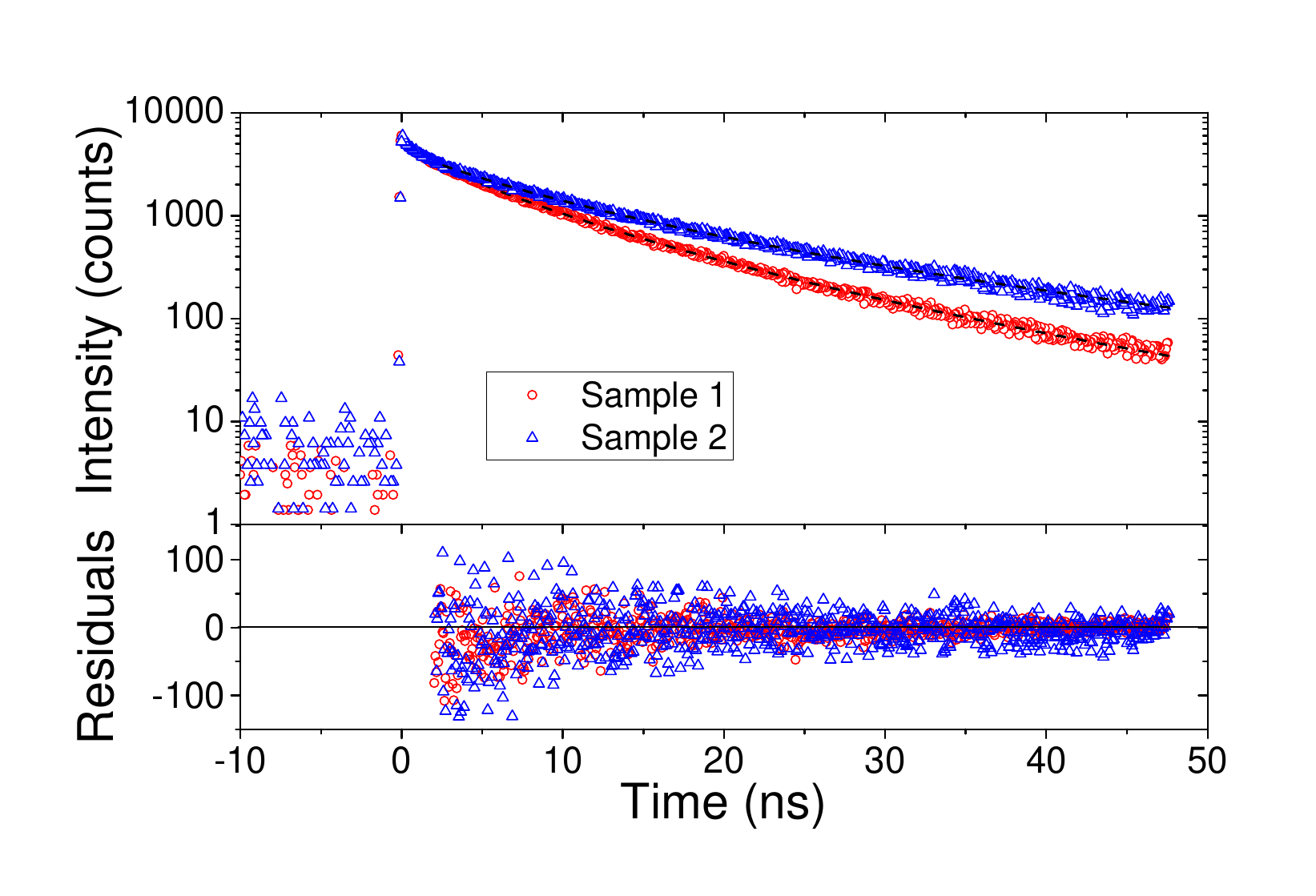}
\caption{(color) Decay curves for quantum dots samples with
different SiO$_2$ layer thicknesses, $z=73$ nm and $z=166$ nm
respectively for sample 1 and 2, measured at an emission energy of
2.08 eV. The decay curves are fitted with a lognormal distribution
of decay rates. Residuals are shown in the bottom panel.
}\label{decaycurvesmirror}
\end{figure}

Measurements of decay rates for two planar samples with different
SiO$_2$ layer thicknesses ($z=73$ nm and $z=166$ nm respectively for
sample 1 and 2) are shown in figure \ref{decaycurvesmirror} for
quantum dots that emit at the peak emission energy of 2.08 eV.
Nonexponential and significantly different decay curves are found
for quantum dots that have different distances to the silver
interface. The quantum dots in sample 1 clearly decay faster than
those in sample 2. The experimental curves are fitted with a
lognormal distribution of decay rates. The residuals shown in the
bottom panel are randomly distributed around a mean value of zero,
signalling a good fit. Indeed the $\chi^2_{red}$ is 0.72 and 1.44
for sample 1 and 2 respectively, close to the ideal value of 1,
signalling that the decay curves are well modeled by a lognormal
distribution of decay rates.

\subsection{Model of decay rates}

Results for the most frequent decay rate for different distances to
the interface are presented in figure \ref{expresults} for two
different emission energies. The most frequent decay rate decreases
with increasing distance to the silver mirror. The measured decay
rate $\gamma_{tot}$ is a sum of radiative $\gamma_{rad}$ and
nonradiative $\gamma_{nrad}$ decay rate,
$\gamma_{tot}=\gamma_{rad}+\gamma_{nrad}$. From Fermi's golden rule
the radiative decay rate is proportional to the projected LDOS
$\rho(\omega,z)$. Therefore, the total decay rate can be expressed
as

\begin{equation}
\gamma_{tot}(\omega,z)= \gamma_{nrad}(\omega)+
\gamma^{hom}_{rad}(\omega)\frac{\rho(\omega,z)}{\rho_{hom}(\omega)}
\end{equation}

Here, $\rho_{hom}(\omega)$ is the LDOS in a homogeneous medium. The
LDOS near an interface has been calculated using a theory developed
by Chance, Prock and Silbey \cite{chance}. As a model an interface
between two semi infinite media has been used, with n$_1=0.27+4.18
i$ (Ag layer) \cite{CRC} and n$_2=1.52$ (SiO2, PMMA and PVA). The
LDOS is calculated for dipoles parallel or perpendicular to the
interface. Our measurements are performed on an ensemble of quantum
dots that are randomly oriented with respect to the interface. This
situation differs from self-assembled dots that are strongly
oriented \cite{johansen}. A decay measurement $f(t)$ for an ensemble
of emitters can be described by the following expression \cite{danz,
koenderink}:

\begin{equation}
f(t)=\frac{I_0}{2 \pi} \int_0^{2\pi} \rm{d}\phi \int_0^{\pi/2}
\rm{d} \theta A(\theta,\phi) ~\gamma(\theta,\phi)~
e^{-\gamma(\theta,\phi)t}~ \sin \theta
\end{equation}

\begin{figure}[!tbp]
\includegraphics[width=4cm]{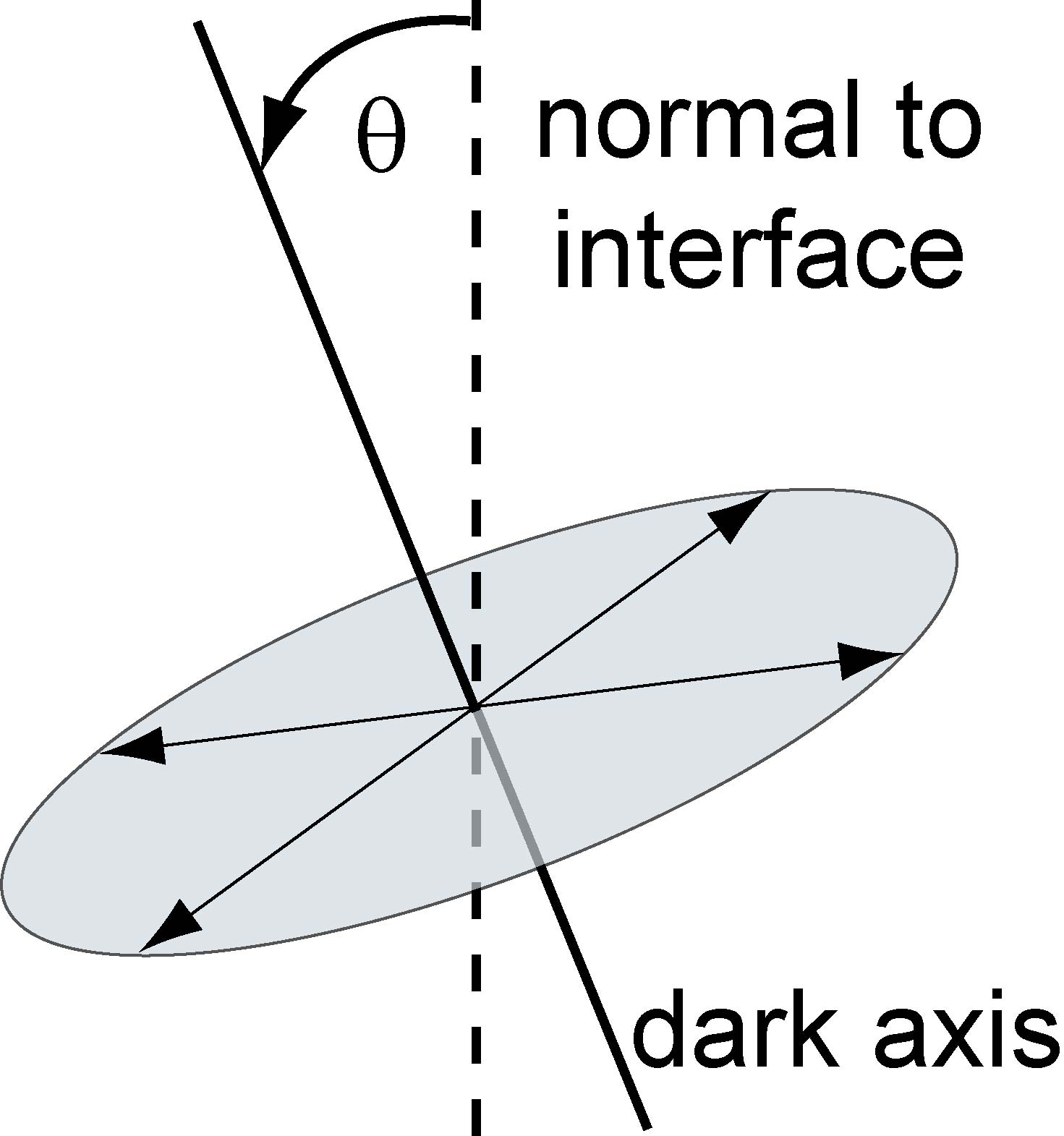}
\caption{The angle $\theta$ is the angle between the dark axis of
the CdSe quantum dot and the normal to the
interface}\label{angledefinition}
\end{figure}

The term $A(\theta,\phi)$ accounts for angle dependence of
absorption, emission and detection. CdSe quantum dots do not have
angle dependent absorption \cite{empedocles2}. Moreover, CdSe
quantum dots are known to have a 2D transition dipole
\cite{empedocles2,efros2} described by a "dark axis" along the
c-axis of the nanocrystal and a "bright plane" perpendicular to this
axis in which the transition dipole can be oriented. Since the
quantum dots have a 2D dipole, the emission is less directional than
if it were a 1D dipole. Because the angle dependence of the emission
and detection plays a small role the factor $A(\theta,\phi)$ can be
safely taken to be independent of $\theta$ and $\phi$. Near an
interface, the decay rate $\gamma$ is no longer dependent on $\phi$
and is given by $\gamma(\theta)=\gamma_{\parallel}
\cos(\theta)^2+\frac{(\gamma_{\parallel}+\gamma_{\bot})}{2}\sin(\theta)^2$
where $\theta$ is the angle between the dark axis of the quantum dot
and the normal to the interface as defined in figure
\ref{angledefinition}. Therefore, carrying out the integral over
$\phi$ results in

\begin{equation}
f(t)=I_0 \int_0^{\pi/2} \Big(\gamma_{\parallel} \cos^2\theta
+\frac{(\gamma_{\parallel}+\gamma_{\bot})}{2}\sin^2\theta \Big)
 ~e^{-(\gamma_{\parallel}
\cos^2\theta+\frac{(\gamma_{\parallel}+\gamma_{\bot})}{2}\sin^2\theta)~t}
~\sin\theta ~{\rm d} \theta
\end{equation}

If $\gamma_{\parallel}=\gamma_{\bot}$ the decay curve shows a single
exponential decay. When $\gamma_{\parallel}$ and $\gamma_{\bot}$
have different values a multi-exponential decay is found. In our
experiment, $\gamma_{\parallel}$ and $\gamma_{\bot}$ only differ by
about at most 10 $\%$. If $f(t)$ is calculated for an intensity
range of 3 decades relevant to our experiment, a single exponential
decay is found to a very high precision with a decay rate given by
$\gamma_{tot}=\frac{1}{3}\gamma_{\bot}+ \frac{2}{3}
\gamma_{\parallel}$. This isotropic decay rate is also used for
experiments with atoms near an interface, where the atom have a
rotating transition dipole moment \cite{barnes}.

\subsection{Discussion}
\begin{figure}[!tbp]
\includegraphics[width=10cm]{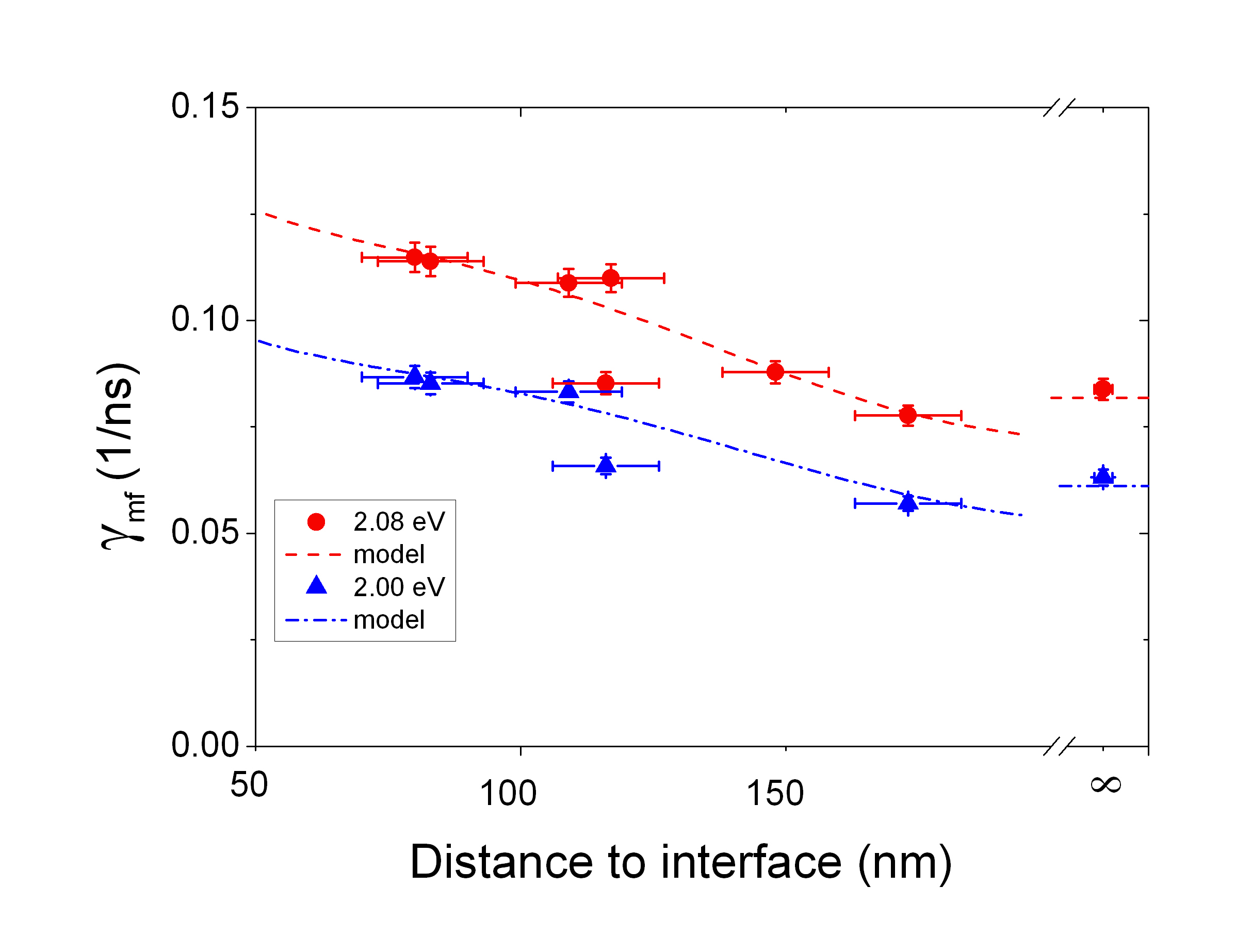}
\caption{Most frequent decay rate $\gamma_{mf}$ versus distance to
the interface for an emission energy of  2.08 eV (red circles) and
2.00 eV (blue triangles). The lines show calculations of the decay
rate using the model developed by Chance, Prock and Silbey
\cite{chance}.}\label{expresults}
\end{figure}

The lines in figure \ref{expresults} show the calculated isotropic
decay rate versus distance to the interface. The calculations are in
very good agreement with the data.

\begin{figure}[!tbp]
\includegraphics[width=10cm]{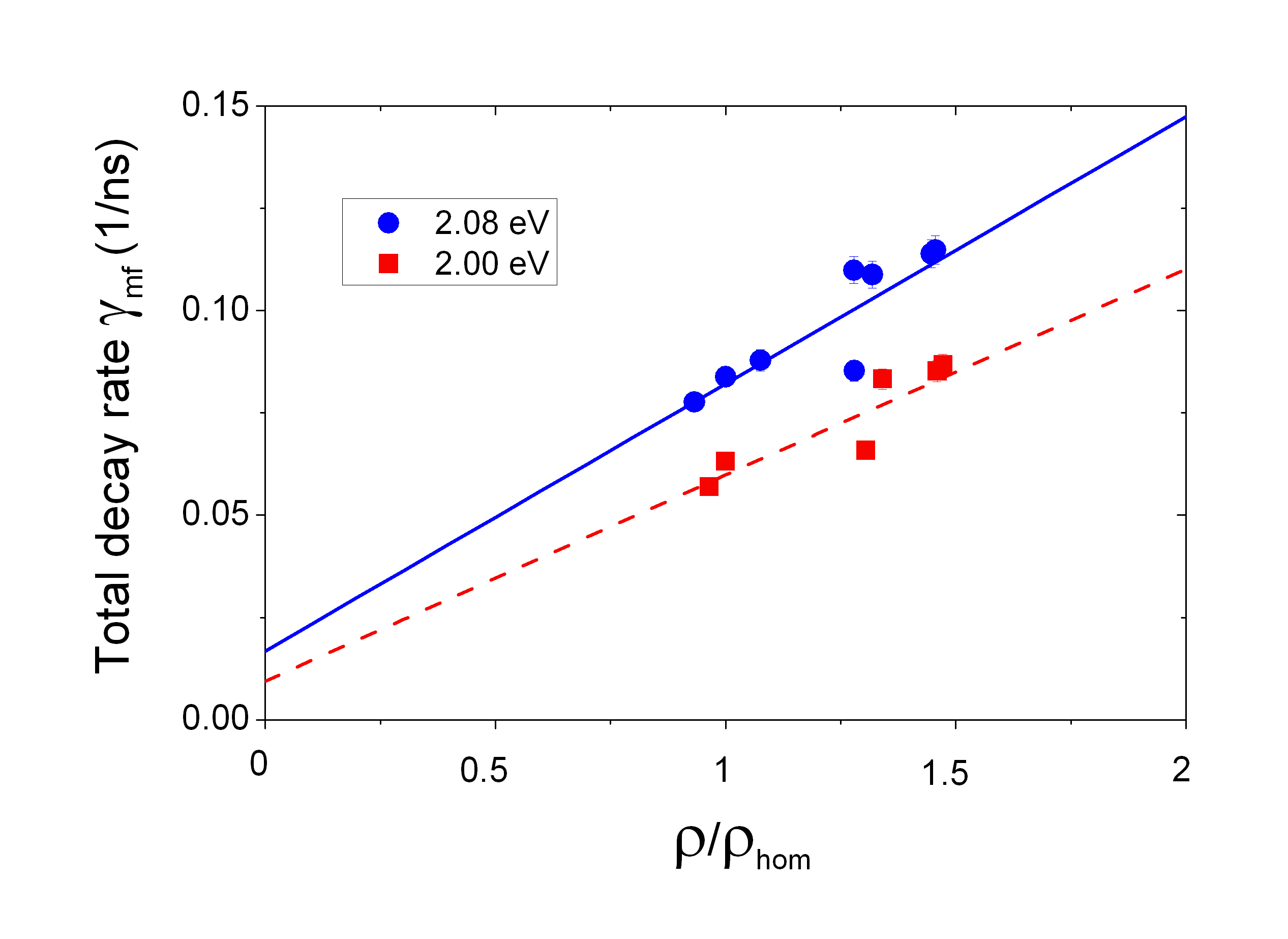}
\caption{The decay rate versus the normalised isotropic LDOS for two
different emission energies. Data are fitted with a linear function
as expected from Fermi's golden rule.} \label{grLDOS}
\end{figure}

By calculating the LDOS for each distance, the distance axis in
figure \ref{expresults} can be converted to an LDOS axis. In figure
\ref{grLDOS} the results are shown for two different emission
frequencies together with a linear fit. Very good agreement between
experiments and theory is observed as expected from Fermi's golden
rule. For an emission energy of 2.08 eV $\gamma_{nrad}=0.017 \pm
0.006$ ns$^{-1}$ and $\gamma_{rad}^{hom}=0.065 \pm 0.005$ ns$^{-1}$
giving a quantum efficiency of 80 $\pm$ 5 \%.

\begin{figure}[!tbp]
\includegraphics[width=10cm]{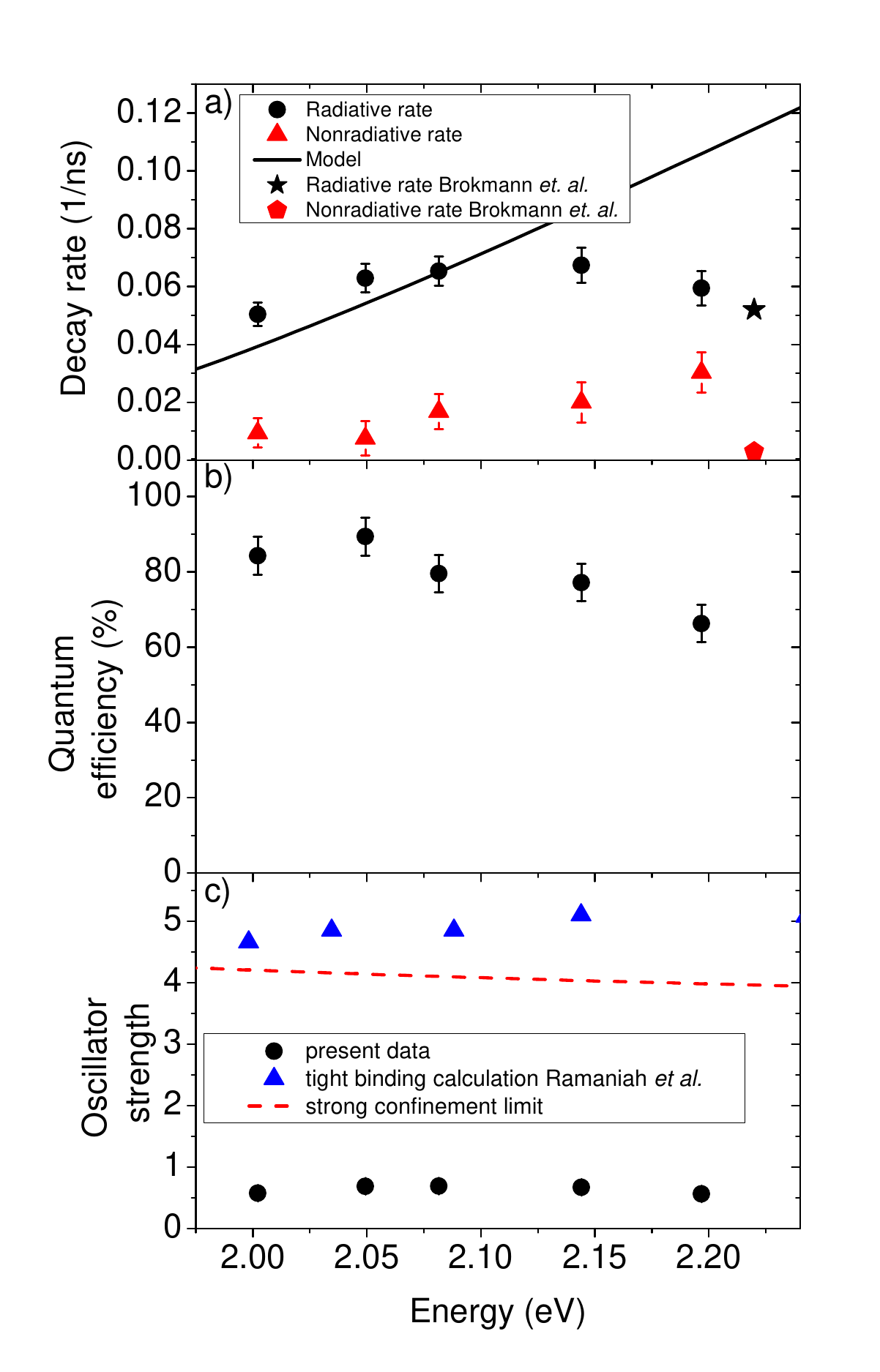}
\caption{ a) Radiative and nonradiative decay rate, determined from
the linear fit in figure \ref{grLDOS} shown for different emission
energies. Values found by Brokmann \textit{et al.} \cite{brokmann}
are also plotted. A model for a multilevel exciton from
\cite{vandriel} is shown as well. b) Quantum efficiency for
different emission energies. c) Oscillator strength for different
emission energies together with a model describing a strongly
confined quantum dot (equation \ref{foscstrong}) and results from
tight binding calculations \cite{ramaniah}.} \label{grsummerygmf}
\end{figure}

In figure \ref{grsummerygmf} a) the nonradiative decay rate
$\gamma_{nrad}$ and homogeneous radiative decay rate
$\gamma_{rad}^{hom}$ are shown as a function of the emission energy
together with the result found by Brokmann \textit{et al.}
\cite{brokmann} for ensembles. The nonradiative decay rate increases
with emission energy or equivalently decreases with quantum dot
size. This is probably due to the fact that for smaller quantum dots
the surface is relatively more important. Since the surface is a
source of nonradiative decay, this decay rate is increased for
smaller quantum dots. An increased nonradiative decay rate for
smaller quantum dots agrees with previous results for CdSe quantum
dots \cite{fan} as well as for epitaxially grown InAs quantum dots
\cite{johansen}. The nonradiative decay rate found by Brokmann
\textit{et al.} for a different batch of quantum dots is lower than
our results. The difference could very well be caused by different
ZnS capping layers since this changes the nonradiative decay
drastically.

The homogeneous radiative decay rate is observed to first increase
and then decrease with emission frequency. The value for radiative
decay rate found by Brokmann \textit{et al.} corresponds very well
to our data. It should be noted that we derived homogeneous
radiative decay from the most frequent decay rate of the
distribution. Since our data agree very well with the decay rate
found using a single exponential model and a much shorter
integration time, this corroborates our choice for the most frequent
decay rate as the parameter that describes the decay curves best.
Our results also validate the choice for the isotropic decay rate
model assumed by Brokmann \textit{et al.}.

Previously the total decay rate (which is the sum of radiative and
nonradiative decay rate) of CdSe colloidal quantum dots was reported
to increase with emission energy \cite{vandriel} as confirmed by our
measurements. A theory was developed for the radiative decay rate as
a function of frequency. For an ideal two level exciton, the
radiative decay rate should be proportional to frequency. If a
multilevel model of the exciton is considered this increase will be
supra-linear. In reference \cite{vandriel} the model for the
excitonic multilevel emitter shows agreement with the total decay
rate data for CdSe quantum dots and excellent agreement for CdTe
dots. The assumption was made that the total decay rate is equal to
the radiative decay rate which is not valid, as can be seen in
figure \ref{grsummerygmf} a). Results for the multilevel exciton
model for radiative decay rate are plotted in figure
\ref{grsummerygmf} a). The model does not fit the data, indicating
that the multilevel exciton model is not a correct model for CdSe.

The quantum efficiency for different emission energies is shown in
figure \ref{grsummerygmf} b). The quantum efficiency is found to be
between 89 and 66 \% depending on emission energy. These values are
significantly higher then the value stated by the supplier Evident,
30-50 \%. This latter value was determined by comparing the emission
intensity to an emitter with known quantum efficiency
\cite{lakowicz}. This method leads however to an underestimation of
the quantum efficiency because it depends on absorption of light.
CdSe quantum dots show strong blinking behavior and quantum dots
that are in the off-state do absorb light, but do not emit. These
quantum dots are considered with an absorption measurement, while
there is no contribution to the emission. This causes an
underestimation of the quantum efficiency in absorption
measurements.

The emission oscillator strength $f_{osc}$ of the transition can be
calculated from the homogeneous radiative decay rate via
\cite{siegman}

\begin{equation}
f_{osc}(\omega)=\frac{6 m_e \epsilon_0 \pi c^3}{q^2 n
\omega^2}\gamma_{rad}^{hom}(\omega)
\end{equation}

where $m_e$ is the electron mass, $\epsilon_0$ is the vacuum
permittivity, c is the speed of light, q is the electron charge and
n is the refractive index of the surrounding material. For an
emission energy of 2.08 eV $f_{osc}=0.69 \pm 0.04$. This is, to our
knowledge, the first experimental value for the oscillator strength
of colloidal quantum dots that is determined by measuring the
photoluminescent emission from quantum dots. Previous qualitative
experiments to determine the relation between oscillator strength
and size of quantum dots used the absorption spectrum of the quantum
dots \cite{leatherdale,schmelz, striolo}. The absorption oscillator
strength is not necessarily equal to the emission oscillator
strength since our measurement is only sensitive to quantum dots
that emit light and are in the on-state, while absorption
measurements probe all quantum dots of the strongly blinking
ensemble, including dots that are in the off-state.

In figure \ref{grsummerygmf} c) the oscillator strength is shown for
different emission energies. The oscillator strength first slightly
increases and then slightly decreases with emission frequency and is
only weakly dependent on energy. Indeed for quantum dots in the
strong confinement regime the oscillator strength is expected to be
only weakly dependent on emission energy since in this regime, the
wavefunctions of electron and hole overlap completely
\cite{brus,kayanuma}. To verify whether this overlap between
electron and hole is indeed unity, the wavefunctions for electron
and hole were calculated using a finite-element method for a simple
effective-mass quantum dot model. The overlap was calculated for a
spherical CdSe quantum dot with a 2 nm ZnS shell. As expected, the
overlap deviated from unity by only 10$^{-4}$ for core radii ranging
from 2 to 4 nanometer.

In the strong confinement limit the oscillator strength is given by
\cite{kayanuma}

\begin{equation}
f_{osc}=\frac{3}{4}\frac{a_B^{*3}}{R^3}\frac{\omega_{bulk}}{\omega_{dot}}
f_{bulk}\frac{\frac{4}{3} \pi R^3}{\frac{1}{2}\frac{\sqrt 3}{2} a^2
c}=\frac{4}{\sqrt3}\pi \frac{a_B^{*3}}{a^2
c}\frac{\omega_{bulk}}{\omega_{dot}} f_{bulk} \label{foscstrong}
\end{equation}

where $f_{osc}$ is the oscillator strength of the quantum dot,
$f_{bulk}$ is the oscillator strength in bulk per chemical CdSe
unit, $a_B^*$ is the exciton Bohr radius, R is the radius of the
quantum dot, $\omega_{bulk}$ is the bulk emission frequency,
$\omega_{dot}$ is the emission frequency of the quantum dot and $a$
and $c$ are the hexagonal lattice constants of CdSe (wurtzite
structure). For $a_B^*=5.4$ nm, $a=0.4302$ nm, $c=0.7014$ nm,
$f_{bulk}=5~10^{-4}$ per chemical CdSe unit \cite{striolo} and
$\omega_{bulk}=2.79~10^{15}$ rad/s the expected curve is shown in
figure \ref{grsummerygmf} c). The calculated values are a factor of
5 larger than the experimentally found values. The oscillator
strength has also been calculated by Ramaniah and Nair
\cite{ramaniah} by a tight binding approach and was found to be 4.9
for a radius of 2.07 nm for spherical CdSe quantum dots. However,
qualitatively in all cases a weak dependence on emission energy is
found that slightly decreases for higher emission energy, in
agreement with our results. Results from absorption measurements
\cite{schmelz, striolo} also find that the oscillator strength is
independent of radius. However, Leatherdale \textit{et al.}
\cite{leatherdale} find a different behavior, seeing a linear
relation between oscillator strength per volume and radius instead
of a cubic dependence.


\section{Conclusion}
In conclusion, we have measured the radiative and nonradiative decay
of CdSe quantum dots by modifying the LDOS in a controlled way. This
allows us to quantitatively determine the oscillator strength and
quantum efficiency versus emission frequency. The nonradiative decay
rate increases with emission energy corresponding to a decrease in
quantum efficiency. The radiative decay rate first increases and
then decreases with energy. This leads to the conclusion that the
increase in total decay rate with energy measured previously is due
to an increasing nonradiative component. The emission oscillator
strength as a function of emission energy is measured with
unprecedented accuracy since for the first time this quantity is
determined without using absorption spectra. The oscillator strength
is weakly size dependent, which is expected in the strong
confinement regime. The oscillator strength is found to be on the
order of 0.7. This is a factor of 5 smaller than expected from
theory and calculated via a tight binding method. The limited
oscillator strength makes the CdSe colloidal quantum dots less
suited for cavity QED experiments. On the other hand, the
quantitative determination of the oscillator strength paves the way
for an ab-initio understanding of spontaneous emission control
\cite{lodahl}.


\section{Acknowledgments}
We thank Oscar Bok, Hans Zeijlemaker and Chris R\'{e}tif for help
with sample preparation and Pedro de Vries and Ad Lagendijk for
helpful discussions. This work was supported by the Stichting
Fundamenteel Onderzoek der Materie (FOM) that is financially
supported by the Nederlandse Organisatie voor Wetenschappelijk
Onderzoek (NWO) and by a VICI fellowship from the Nederlandse
Organisatie voor Wetenschappelijk Onderzoek (NWO) to W.L.V.

\newpage
\section{Appendix: Conclusions for relative width of the distribution}
In this work results are presented of the effect of modified LDOS on
the most frequent decay rate. This most frequent decay rate is found
by fitting a lognormal distribution of decay rates to the
experimental decay curves. The other independent fitting parameter
in this fit is the relative width of the lognormal distribution. In
this appendix results for the relative width are presented.

\begin{figure}[!tbp]
\includegraphics[width=10cm]{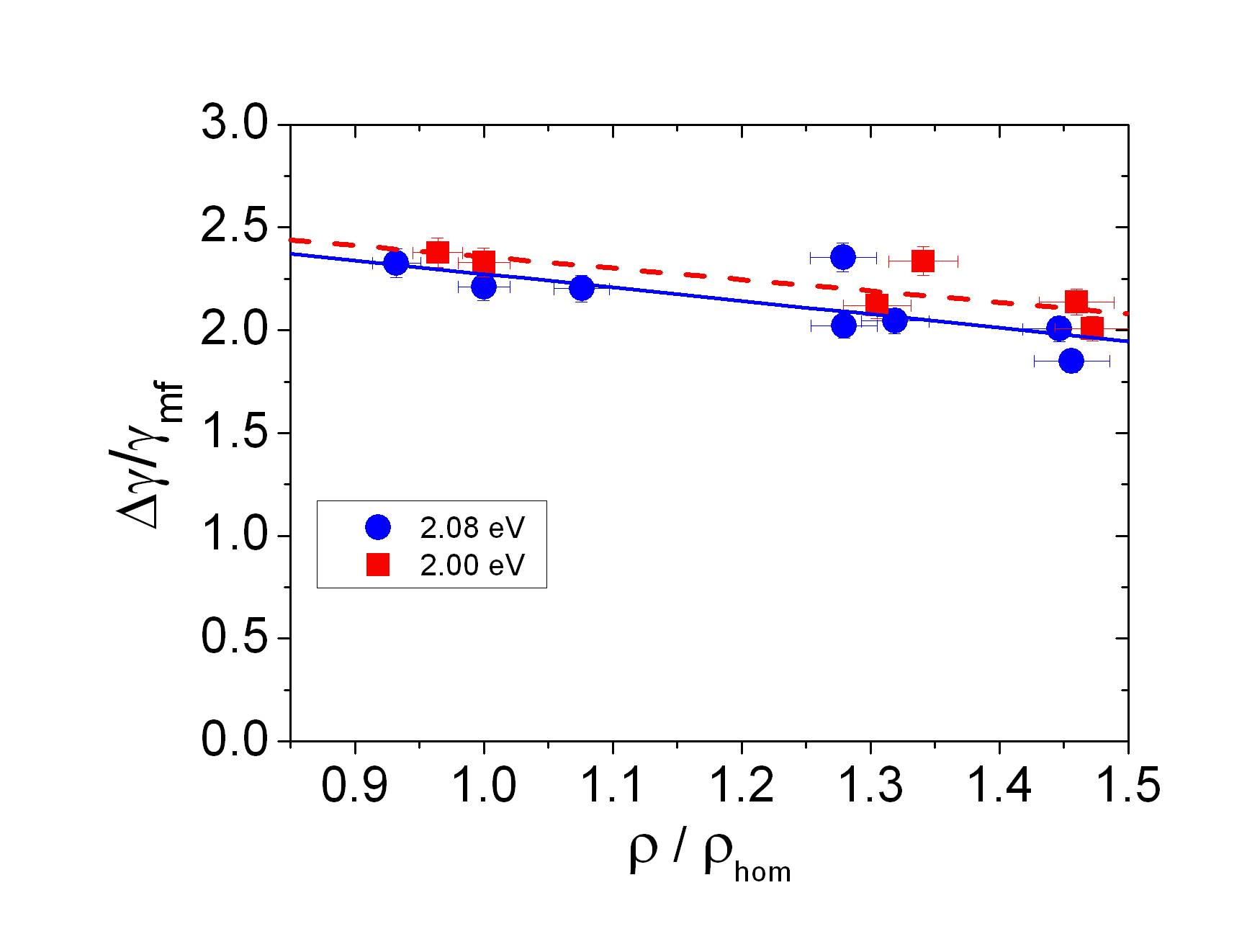}
\caption{Relative width of the lognormal distribution versus LDOS
for emission energies of 2.08 and 2.00 eV. The lines are linear fits
of the data.}\label{relwidthvsldos}
\end{figure}

In figure \ref{relwidthvsldos} the relative width, defined as
$\frac{\Delta\gamma}{\gamma_{mf}}$, is plotted versus normalised
local density of states for emission energies of 2.08 and 2.00 eV.
For increasing LDOS the relative width decreases linearly.
Increasing the LDOS effectively increases the quantum efficiency
because the radiative decay rate is increased while the nonradiative
decay rate is constant. For increasing quantum efficiency the
distribution in decay rates gets narrower, giving a strong
indication that the width of the distribution is determined by the
nonradiative decay rate confirming the proposition by Fisher
\textit{et al.} \cite{fisher}.

\begin{figure}[!tbp]
\includegraphics[width=10cm]{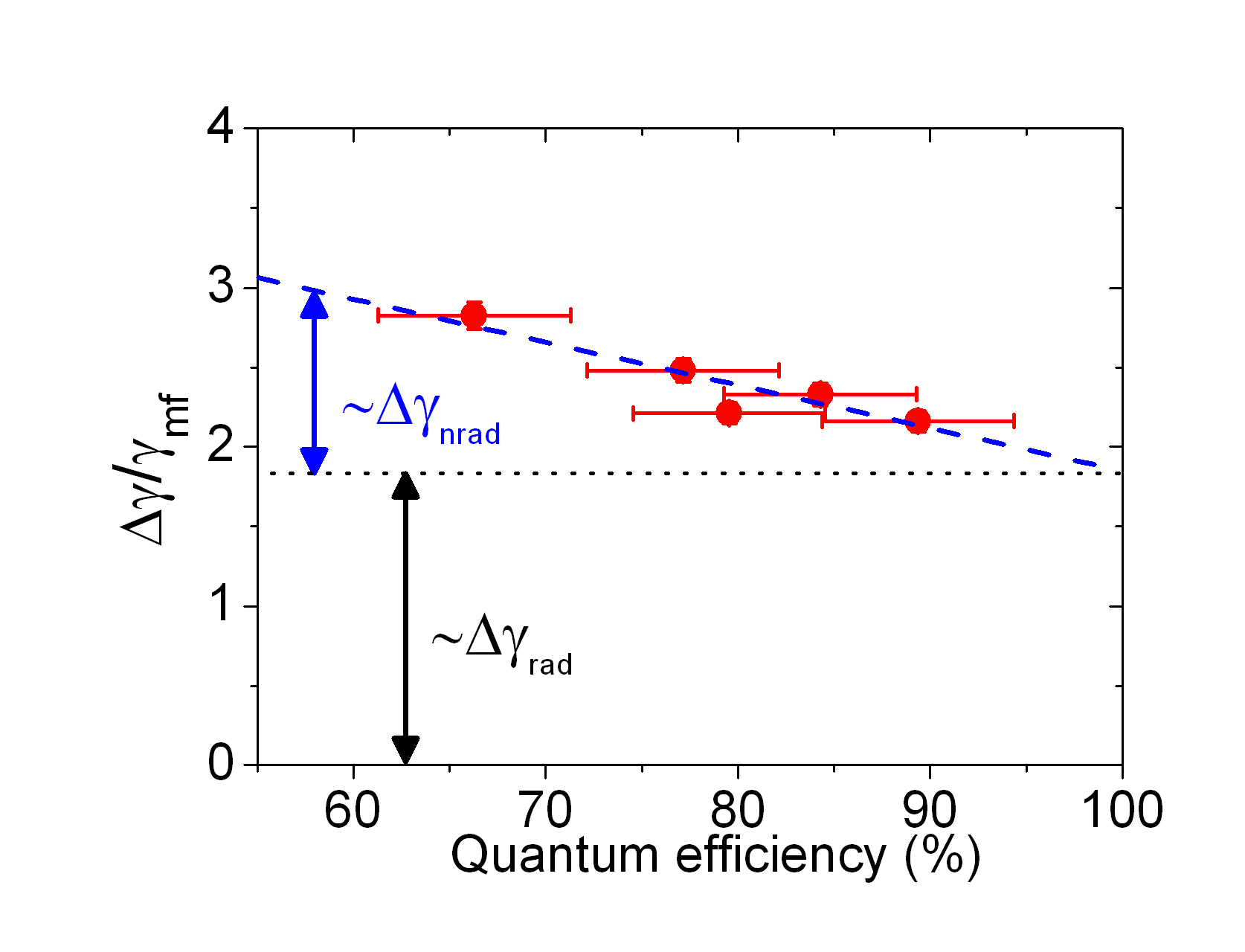}
\caption{Measurements of relative width for a homogeneous system
(LDOS=1) plotted versus the extracted quantum efficiency (see figure
\ref{grsummerygmf} b) together with a linear
fit.}\label{relwidthvsqe}
\end{figure}

In figure \ref{relwidthvsqe} the relative width measured in the
homogeneous environment with LDOS = 1 is plotted versus the
extracted quantum efficiency for each emission energy. The same
trend is found. For increasing quantum efficiency the relative width
of the distribution decreases linearly. When the quantum efficiency
is 100 \%, the decay rate is purely radiative. If the width in the
distribution of decay rates is only caused by the nonradiative rate,
the width should be zero at 100 \% efficiency. This is not the case,
indicating that there is a distribution in radiative decay rate as
well. Vall\'{e}e \textit{et al.} \cite{vallee} have also found
distributions of decay rates for single dye in polymer and attribute
this to local density variations in the surrounding polymer matrix
causing a distribution in radiative decay rate.

In conclusion, our data shows that there is both a distribution in
nonradiative and radiative decay rate that cause the distribution in
total decay rate.

\end{document}